# RAPID CONNECTIONIST SPEAKER ADAPTION

Michael Witbrock[1,2] and Patrick Haffner[2]

[1]School of Computer Science, Carnegie Mellon University, Pittsburgh, PA 15213-3890 USA

[2]Centre National des Etudes de Télécommunications, Route de Trégastel, Lannion, 22301 France

## ABSTRACT

We present SVCnet, a system for modelling speaker variability. Encoder Neural Networks specialized for each speech sound produce low dimensionality models of acoustical variation, and these models are further combined into an overall model of voice variability. A training procedure is described which minimizes the dependence of this model on which sounds have been uttered. Using the trained model (SVCnet) and a brief, unconstrained sample of a new speaker's voice, the system produces a Speaker Voice Code that can be used to adapt a recognition system to the new speaker without retraining. A system which combines SVCnet with an MS-TDNN recognizer is described.

## INTRODUCTION

Speaker dependent speech recognition systems perform substantially better than speaker independent ones. In an attempt to narrow this performance gap, a great deal of research has been done to develop systems that adapt to the voice of a new user.

These systems are not yet an adequate solution to the problem of voice variability since they do not match some important features of human speaker adaptation: humans can improve their recognition performance for a new speaker in real time, and after only a few words have been heard [ll. This level of adaptation ability is not only desirable in terms of matching human performance, it may be essential if we are to get good performance on tasks, such as telephone enquiry handling, for which gathering an extensive adaptation set is impossible.

This paper describes SVCnet— a system that learns to identify where a speaker's voice lies in a space of possible voices. The speaker voice code output by SVCnet has the following when used 00 control adaptation:

- It can be computed using a completely unconstrained sample of the speaker's voice.
- It can be computed during recognition.
- It can be used by an appropriately recognizer to adapt to a new speaker's voice without retraining.
- It can be computed rapidly, reaching a stable value after only a few words have been spoken.

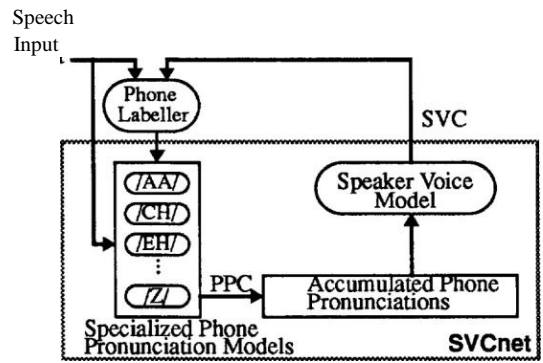

**Fig. 1. SVCnet's high-level architecture.**

SVCnet differs from other speaker adaptation systems (e.g. [2,3]) in that during training, it learns to model the variation in the voices of a great many speakers, and can therefore identify an appropriate adaptation during recognition, rather than having to derive one from the incoming speech. Its architecture is based on the observation that neural networks trained to reproduce their inputs form a compact representation of those inputs in their hidden units.

We present a description of the architecture of the SVCnet system and then show how it is trained and used. Examples of the speaker voice codes developed by the system are described, followed by an example of their use to adapt an MS-TDNN Speech Recognizer [4].

## THE SVCNET SYSTEM

The SVCnet system works with a speech recognition system 10 produce a speaker voice model that is, in turn, used to improve the recognizer's performance. The general system architecture is shown in Fig. 1. The system builds a description of the speaker's voice hierarchically, first describing how phones (or other sub-word units) pronounced by that speaker differ from those of other speakers, and then combining these descriptions (Phone

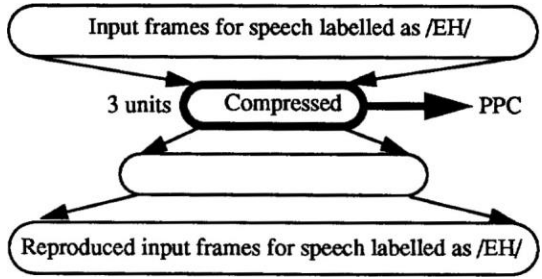

**Fig. 2. One of the networks trained to model the variation in a phone (this one models variation in /EH/). These networks construct the Phone Pronunciation Code.**

Pronunciation Codes or PPCs) over time to produce an overall Speaker Voice Code (SVC).

The labels output by the speech recognizer are used to direct the incoming frames of speech to a network specialized for producing a description of the sound, and the SVC is used by the recognizer to improve its labelling performance.

**Phone Pronunciation Codes**

To model the pronunciation differences that make up voice differences, we make use of the ability of neural networks with a narrow hidden layer (a bottleneck) to form a reduced dimensionality encoding of their input. The elementary component from which the speaker modelling system is built up is such an encoder network trained, using backpropagation, to model the variation in the pronunciation of a particular sound by different speakers [Fig. 2.]. Each network tries to reconstruct its input after passing it through a

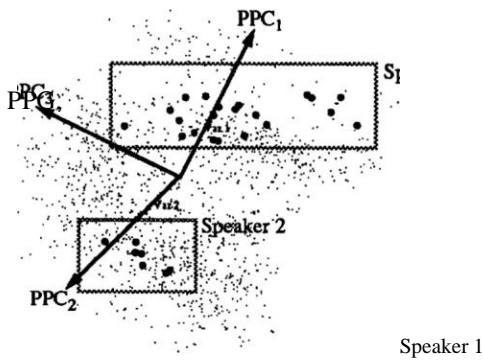

Speaker 1

**Fig. 3. Phone Pronunciation codes for the phoneme [EH/ for a large number of RMSpell speakers. Codes from two speakers are highlighted. The axes have been rotated to accentuate the difference between speakers.**

narrow bottleneck and, in doing so, sets the activities of the units in this bottleneck to values that constitute a model of the variation among the examples of the sound, or, approximately, to the variation in speakers' pronunciation of the sound. Fig. 3. shows the Phone Pronunciation Codes (PPC) formed for various frames from the phoneme [EH/ for a large number of different speakers, two of which are highlighted

In the present system, words or phones are further divided into states using a DP alignment procedure. This division increases the number of modelled sounds and reduces the variation in the PPC resulting from other sources than speaker differences.

**The Speaker Voice code**

Our aim is to form a representation of the speakers' voice (the SVC) that is independent of the words used to form it and that can be developed after only a few words have been spoken. So far, we have described a way of producing representations of the way the sounds in these words have been pronounced (the PPCs). Another bottleneck network shown in Fig. 4. is trained to produce the SVC as follows: appropriate PPCs are formed for all the frames of speech available from a particular training speaker. Multiple PPCs for a sound are averaged, yielding a single PPC for each sound uttered by the speaker. This collection of phone pronunciation codes is the target that the network is trained to reproduce on its output units (no error is backpropagated from units corresponding to sounds not uttered by that speaker). Once the target has been determined in this way, the system returns to the beginning of the utterance, and steps through it, accumulating PPCs on the inputs to the network as the corresponding sounds are pronounced (this time, a running average is kept for repeated PPCs). Weight updates are performed after each input presentation. Training is performed in this way until the whole utterance is complete and then repeated in the same way for the next speaker. To summarize, the speaker modelling network is trained to produce the complete set of PPCs from the increasingly large subset of those models available as more

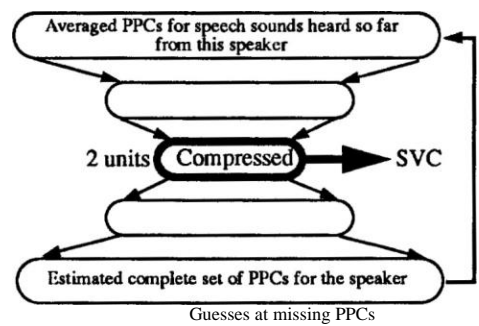

Guesses at missing PPCs

**Fig. 4. This network does pattern completion for PPCs. While doing so, it forms a Speaker Voice Code that is largely independent of the sounds heard.**

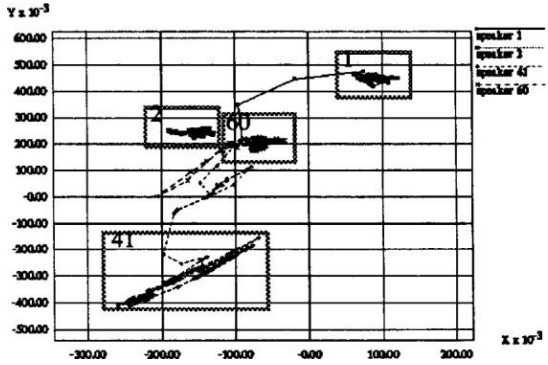

**Fig. 5. SVC formation over time for four speakers from the RMSpell database. Four points on a trajectory represent one word. SVCs reach a relatively stable point in speaker space after a few words of speech**

speech is heard; it guesses how all speech from the speaker will sound on the basis of the subset it has already heard.

After some research, a variation in this training scheme was adopted that significantly reduced the time taken to form a stable speaker model. Instead of leaving the inputs for not-yet-heard states at zero, output unit activities for these states are fed back to the input. The network combines its increasingly accurate guesses of what the pronunciation models will be with the incoming actual pronunciation models to form a stable speaker code after about 4 words have been heard. Examples of the formation of S VCs over time are given for several speakers in Fig. 5. The position of the SVC is reasonably stable after about 3 words (in this case, three letters) have been spoken.

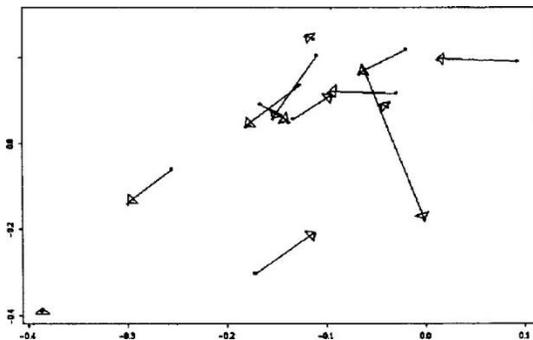

**Fig. 6. The arrows in this figure show the difference between SVCs formed from two different sets of 7 utterances from a subset of the speakers in the RMSpe11 database.**

---

[1] This DARPA database consists of speech from 120 speakers from the U.S.A, each spelling 15 words out loud.

| Acoustic | State | Word | % Errors |
|---|---|---|---|
| ✗ | ✗ | ✗ | 1.77 |
| ✗ | ✗ | ✓ | 1.62 |
| ✗ | ✓ | ✗ | 1.74 |
| ✗ | ✓ | ✓ | 1.47 |
| ✓ | ✗ | ✗ | 1.17 |
| ✓ | ✗ | ✓ | 1.14 |
| ✓ | ✓ | ✗ | 1.11 |
| ✓ | ✓ | ✓ | 0.99 |

*Table 1. Error rates with SVC available to various levels of the MS-TDNN during testing.*

Fig. 6. shows that the SVC for most speakers is relatively independent of the content of the utterances from which it is generated. Each arrow corresponds to one speaker from the RMSpell database[1] with the tail lying at the average value of the SVC generated from the 1st 7 words spelled by that speaker, and the head of the arrow pointing to the SVC generated by the second 7 words spelled by the speaker.

## APPLYING SVCNET

We view speaker adaptation as the process of recognizing — as opposed to learning — where in speaker space a new speaker lies, and of using this information to bias the recognition mechanism so that it will perform correctly for that speaker.

We have done some very preliminary experiments in using the SVCnet system to adapt an MS-TDNN (described in [4]) designed to recognize telephone quality French digits. There were 3540 digits in the training set and 3335 digits, from different speakers, in the testing set. Each speaker uttered, on average, 9 of the 10 French digits. The SVCs for these experiments were generated off-line using word labels generated using an accurate HMM recognizer and were averaged over the entire utterance for each speaker.

The SVC was made available as additional input to the MS-TDNN using two additional input units, fully connected to two additional hidden units, that were in turn connected to every unit in the MS-TDNN. While information from the SVC was available to all units in the MS-TDNN during training, during testing it could be replaced, for a given layer, by its average value across speakers.

### Performance results on digit recognition

Table 1. gives the performance of the MS-TDNN with the code available (✓) or not available (✗) to each of the three

hidden layers (acoustic, state, and word) of the MS-DNN recognizer.

There is a general trend in the data that the more places in the network the speaker code is available, the better it does. The most substantial improvement occurs when the Speaker Voice Code biases the 1st hidden layer, which is responsible for identification of acoustic features. At this level, the MS-TDNN could use the SVC to effect an acoustic normalization. The second layer combines acoustic features into state scores, and the speaker model can influence the relative importance of these features. The third layer combines state scores into word scores, so the influence of the SVC here represents, approximately, a transition penalty for a state.

The best performance for a similar MS-TDNN system trained with ordinary non-speaker-dependent biases was reached at 1.1% error. This relatively good result, which seems surprising at first, is explained by the fact that the ordinary MSTDNN has available to it a considerable span (>100ms) of speech context from which it can derive an approximate speaker model. Since, in our system, the MS-TDNN has the speaker code provided to it by the SVCnet system, it suffers when it is deprived of this speaker information. When the speaker code is available to the whole system, as it expects, it does somewhat better that the ordinary MSTDNN, and can be expected to do better yet when a global, simultaneous optimization of the speaker-modelling and recognition component of the system is done.

**Speed of SVC formation**

To confirm that the SVC can be formed rapidly and can be formed from a non-tested sample of speech, the recognizer was tested on the last four digits from each speaker (1532 digits total). The speaker voice codes used were derived from either the same four digits, or from the first four, different, digits from the same speaker.

| Source of SVC | % Errors |
|---|---|
| No speaker voice code | 1.50 |
| First 4 (different) digits | 0.85 |
| Last 4 (same) digits | 0.78 |

*Table 2. Performance of the system for the last 4 digits from a speaker, with the SVC derived from different subsets of the digits spoken.*

This table shows that, again, the system does better when it has the speaker code available than when it does not, and that its performance depends little on whether the speaker code is extracted from the same digits it is trying to recognize or from different ones.

## CONCLUSION

We have introduced a system that learns to model the variation found between speakers, and that can use this model and a short, unconstrained sample of speech to identify where a new speaker lies in speaker space. This information, in the form of a Speaker Voice Code, can easily be provided as additional input to a neural network speech recogniser, allowing it to adapt its performance to a new speaker without any retraining.

In future work, we will apply the SVCnet adaptation system to a recognizer for the RMSpell database, a task with more potential to benefit from adaptation. We will attempt to improve the SVC so that it shows less variation within a speaker and more between speakers. We will also investigate the degree to which the speaker space modelled by SVCnet corresponds to the type of speaker variation perceived by humans.

## ACKNOWLEGEMENTS

The authors acknowledge the support of CNET, DARPA, the National Science Foundation, and would like to thank Scott Fahlman, Yann le Cun, Laurent Miclet, Alex Waibel and Members of the CNET Speech Recognition Research Department for valuable technical discussions.